\documentclass[pra, aps,
twocolumn,
amsmath,amssymb
,floatfix
]{revtex4-1}

\usepackage{physics}
\usepackage{epsf}
\usepackage{float}
\usepackage{graphicx}
\usepackage{amsmath}
\usepackage[usenames]{color}
\usepackage{float}
\usepackage{bm}
\usepackage{relsize}
\usepackage[normalem]{ulem}

\def \be {\begin{equation}}
\def \ee {\end{equation}}
\def \bea {\begin{align}}
\def \eea {\end{align}}
\def \p {\partial}
\def \BEA {\begin{eqnarray}}
\def \EEA {\end{eqnarray}}
\def \BC {\begin{cases}}
	\def \EC {\end{cases}}
\usepackage{siunitx}
\usepackage{tabularx}
\usepackage{xcolor}
\usepackage[version=4]{mhchem}
\usepackage{ctable}

\usepackage[unicode=true,colorlinks=true,citecolor=blue,urlcolor=blue]{hyperref}

\def \be {\begin{equation}}
\def \ee {\end{equation}}
\def \bea {\begin{align}}
\def \eea {\end{align}}
\def \p {\partial}
\def\bee{\begin{eqnarray}}
\def\eee{\end{eqnarray}}
\def \BC {\begin{cases}}
	\def \EC {\end{cases}}

\begin{document}

\title{Thermoelectric and viscous contributions to the hydrodynamic ratchet effect
}

\author{ S.~O. Potashin$^1$, L.~E. Golub$^{2}$, V.~Yu. Kachorovskii$^{1}$}

\affiliation{$^1$Ioffe Institute, 194021 St. Petersburg, Russia}
\affiliation{$^2$Terahertz Center, University of Regensburg, 93040 Regensburg, Germany}

\

%\date{\today, v.3}

\begin{abstract} 
	We study  thermoelectric and viscous  contributions  to  the   ratchet effect, i.e.   radiation-induced generation of the direct electric current, $J_{\rm rat},$  in asymmetric  dual-grating gate structure  without inversion center.  Previously [E.Mönch et al, Phys. Rev. B {\bf 105}, 045404 (2022)], it was demonstrated that frequency dependence  of the  $J_{\rm rat}$ is essentially different within   hydrodynamic (HD)  and  drift-diffusion (DD) regimes  of the electron  transport:  $ J_{\rm rat}^{\rm HD} \propto 1/\omega^6  $ and    $ J_{\rm rat}^{\rm DD} \propto 1/\omega^2 $  for $\omega \to \infty.$  Here we  analyze  previously neglected thermoelectric contribution and find that it  yields  high-frequency asymptotic $1/\omega^2$ even in the HD regime and can change sign of the response.    Account of the  finite viscosity of the electron liquid  yields contribution which  scales at high frequency as   $1/\omega^4.$   We also find plasmonic resonances  in the   $J_{\rm rat},$ and demonstrate that asymmetry of the structure allows for excitation of the so-called directional travelling plasmons.  
 %The obtained results  are used to analyze conditions needed for realization  of th HD regime in the optical experiments 
  % in response to external terahertz radiation  with  asymmetric density modulation  in bilayer graphene, where inversion symmetry is broken by an asymmetric dual-grating gate potential. As a central result, we demonstrate that at high temperature, T=150 K, the ratchet current decreases at high frequencies as $∝1/ω2$, while at low temperature, T=4.2 K, the frequency dependence becomes much stronger ∝1/ω6. The developed theory shows that the frequency dependence of the ratchet current is very sensitive to the ratio of the electron-impurity and electron-electron scattering rates. The theory predicts that the dependence $1/ω6$ is realized in the hydrodynamic regime, when electron-electron scattering dominates, while $1/ω2$ is specific for the drift-diffusion approximation. Therefore, our experimental observation of a very strong frequency dependence reveals the emergence of the hydrodynamic regime.
\end{abstract}

\maketitle

\section{Introduction}
\label{intro}

%\begin{widetext}

One of the most general and  fascinating phenomena in optoelectronics is
the ratchet effect---the generation of a dc electric current
in response to an ac electric field in systems with broken
inversion symmetry (for reviews see, e.g., Refs. \cite{Linke2002,Reimann2002,Haenggi2009,Ivchenko2011,Denisov2014,Cubero2016,Budkin2016a,Reichhardt2017,Ganichev2017}).  In particular,
this general definition can be used for artificial structures aimed to  modulate periodically the electron density in the 2D channel:    long periodic grating
gate structures with an asymmetric configuration of
gate electrodes, e.g., dual-grating top gate (DGG) structures
\cite{Olbrich2011,Otsuji2013,BoubangaTombet2014,Faltermeier2015,Olbrich2016}. 
The ratchet effect was treated theoretically and observed
experimentally in various low dimensional structures
\cite{Olbrich2011,Otsuji2013,BoubangaTombet2014,Faltermeier2015,Olbrich2016,Faltermeier2017,Olbrich2009a,Popov2011,Kannan2011,Nalitov2012,Kannan2012,Kurita2014,Rozhansky2015,Bellucci2016,Fateev2017,Faltermeier2018,Rupper2018,Yu2018,Hubmann2020,DelgadoNotario2020,Sai2021,Yahniuk2024,Hild2024}, so that the ratchet current measurements
can already be considered as a standard  measurement tool. 
It has recently been shown that the ratchet effect can also be used to observe the  transition of an electron system  into hydrodynamic (HD) regime.        
Specifically,  it was demonstrated in Ref.~\cite{Moench2022}, both theoretically and experimentally,
that the dc response of a DGG, based on bilayer
graphene, has very   strikingly different frequency dependencies: $1/\omega^6$ in the HD regime and $1/\omega^2$ within
the so-called  drift-diffusion (DD) approximation. An analytical
formula which describes the transition between both
regimes was derived  which was in a good agreement with obtained experimental data.  The key statement  of this work, namely, HD-like behavior for liquid helium temperature  was  also confirmed  by  publication       \cite{Moench2023} focused on measurement of ratchet current in the   Shubnikov de Haas regime. As was argued in  Ref.~\cite{Moench2023},   the   experimentally observed  strong suppression of the second harmonic of the cyclotron resonance  indicates presence of    
 fast electron-electron collisions  that drive the electron system into HD regime. 
 
 The results obtained in Refs.~\cite{Moench2022,Moench2023} open a wide avenue for search  of the  HD regime  in  the  optical experiments. This problem has a long history and now the   
electronic fluid dynamics is one of the extremely actively developing areas of condensed matter physics (for review, see, e.g., Refs.~\cite{Narozhny2017, Lucas2018,Narozhny2019, Polini2020}). Although the pioneering works~\cite{Gurzhi1963, Gurzhi1965, Gurzhi1968, Jong1995} on hydrodynamic electron and phonon transport
%were written
have been done a very long time ago, the topic did not generate much interest until recently.
%The trigger of booming  interest in hydrodynamic transport was
The interest on hydrodynamic transport was triggered by the fabrication of ultraclean ballistic structures, primarily based on one-dimensional and two-dimensional carbon materials. Convincing manifestations of hydrodynamic behavior in the different transport regimes have been demonstrated in a number of recent experiments ~\cite{Bandurin2016, Crossno2016, Moll2016, Ghahari2016, Kumar2017, Bandurin2018b, Braem2018, Jaoui2018, Gooth2018, Berdyugin2019, Gallagher2019, Sulpizio2019, Ella2019, Ku2020, Raichev2020, Gusev2020, Geurs2020, Kim2020, Vool2020, Gupta2021, Gusev2021, Zhang2021, Jaoui2021}. Moreover, literally in recent years, it has been possible to experimentally
visualize the hydrodynamic flow  in ballistic 2D systems by  using various nanoimaging techniques~\cite{Braem2018, Sulpizio2019,Ella2019, Ku2020, Vool2020}.     

All previous publications  searching the HD regime  were focused on   measurements of  dc viscous   transport.   However,  the approach based on optical experiments 
 has a number of advantages.
 %as compared to the one based on  the study of the viscous  dc  transport.
 %, e.g. see in Refs.~\cite{Jong1995, %Bandurin2016}. .   
% $$$$ $$$$ $$$$
% A few comments need to be made on the temperature dependence of the effect and on    the differences   of  our approach with previous analysis of viscous  flow   of the electron liquid.
% The temperature range in which the observation of the hydrodynamic regime is possible is limited in both from above and from below. The lower limitation is caused by the decrease of the electron-electron scattering rate with the temperature, whereas the scattering rate on impurities is almost temperature-independent. %\sout{is because the rate of electron-electron scattering decreases with temperature, and the rate of scattering by impurities is practically independent of temperature.}
% Therefore, at sufficiently low temperatures, electron-electron scattering is "turned off" and the system goes into the drift-diffusion regime. On the other hand, with increasing temperature, sooner or later, strong scattering by phonons comes into play, which in the context of the problem under study is like scattering by impurities. Accordingly, there is an upper limit on the temperature. 
%Importantly, as one can see from our data, the lower temperature limit for realization of the HD regime in our case is smaller as compared to realization of the viscous flow of the electron fluid studied experimentally, e.g. see in Refs.~\cite{Jong1995, Bandurin2016}. 
Indeed, one of the hallmarks of the viscous dc flow  %Poiseuille flow 
is the Gurzhi effect predicted in Refs.~\cite{Gurzhi1963, Gurzhi1965, Gurzhi1968}. Starting from its first experimental observation in Ref.~\cite{Jong1995},  this effect is considered as one of the most convincing arguments in favor of viscous transport. The Gurzhi effect is observed in a system of finite transverse (with respect to electron flow)  width $d$ under the assumptions $l_{\rm ee} \ll d \ll L_{\rm G},$ where $l_{\rm ee}$ is the electron-electron collision length, $L_{\rm G} =\sqrt{l l_{\rm ee}}$ is the so called Gurzhi length, and $l$ is the  momentum relaxation length limited by impurity and phonon  scattering. The inequality $d \ll L_{\rm G} $ is not easy to satisfy in a sufficiently wide sample. That is why for the observation of viscous transport it is necessary to use narrow-channel samples with ultrahigh mobility. Also, for the observation of negative non-local resistance, see Ref.~\cite{Bandurin2016}, the size of the viscosity-induced whirlpools responsible for viscous back flow (this size is in the order of $L_{\rm G}$) was in the order of the size between contacts probing a negative voltage drop. If one uses thin wires or narrow strips for the observation of the Gurzhi effect, the second inequality, $l_{\rm ee} \ll d$ can be satisfied only at sufficiently high temperatures. By contrast, in the optical experiments  one can  study a bulk effects which do not disappear with increasing system size and distance between contacts. For the observation of the HD transport, one only need the condition $l_{\rm ee} \ll l,$ which is independent of the system size.

Moreover, the HD regime is not necessarily viscous. The key property of the HD regime (as compared to the DD one) is the presence of only three collective variables (local temperature, concentration and drift velocity), which completely characterize the system, in contrast to the DD regime, where the distribution function is not reduced to a hydrodynamic ansatz depending, in the general case, on an infinite number of variables. Hence, HD  regime can be realized for the case when the viscous contribution to the resistivity is small: 
\be \eta q^2 \ll \gamma, 
\label{eta}
\ee   
where  $\eta= v_{\rm F}^2 \tau_{\rm ee}/4 $ is the electron viscosity, $\gamma= v_{\rm F}/l =1/\tau,$ $\tau $ is the momentum relaxation time,  and $q$ is the characteristic inverse scale of the inhomogeneity of the problem (inequality \eqref{eta} is equivalent to condition  $q L_{\rm G} \ll 1$).  It is worth also noting, that  the latter condition is always satisfied in the limit of the ideal liquid, $\tau_{\rm ee} =  0.$        

On the other hand,   an additional possibility to observe the HD regime  appears  in the non-linear excitation regime discussed in  Refs.~\cite{Moench2022,Moench2023}.  In such a regime, in contrast to the linear one, the difference between the distribution functions in the DD and the HD regimes is very strong and causes currents, which are strongly different even if one neglects viscosity. This allows one to distinguish between the DD and the HD regimes  even when the condition Eq.~\eqref{eta} is satisfied. In particular, this condition was assumed to be satisfied in Ref.~\cite{Moench2022,Moench2023}, where the viscosity contribution was neglected.

In this paper, we 
focus on the  thermoelectric contribution to $J_{\rm rat}$  and also
take into account effect  of finite  viscosity still assuming that condition Eq.~\eqref{eta} is satisfied.  Thermoelectric contribution was   neglected  in Ref.~\cite{Moench2022}
%.  As for thermoelectric  effects, they were neglected 
by   assuming that (see discussion in Ref.~\cite{Nalitov2012})  
\be 
\Gamma= \frac{\tau}{\tau_{\rm T}} =   \frac{3 \tau }{ \pi^2 \tau_{\rm ph}} \frac{\mu}{T}\gg 1 , 
\label{thermo-cond}
\ee  
where $T$ is the temperature, $\tau_{\rm T} = \tau_{\rm ph} \mathcal C/N = \pi ^2 T \tau_{\rm  ph} / (3 \mu)$ 
is  the temperature relaxation time, $ \mathcal C=\nu T \pi^2/3$  and $\nu$ are, respectively, the heat capacitance and density of states  of the 2D Fermi gas, $\mu$ is the Fermi energy, and $\tau_{\rm ph}$  is the phonon scattering rate.
We  demonstrate that  previously neglected thermoelectric contribution  yields  high-frequency asymptotic $1/\omega^2$ even in the deep HD regime ($\tau_{\rm ee}=0$)   and can change sign of the response  with decreasing $\Gamma$.    Account of the  finite viscosity of the electrton liquid  yields contribution scaling at large frequency as   $1/\omega^4.$  We also find plasmonic resonances  in   $J_{\rm rat},$ and demonstrate that asymmetry of the structure allows for excitation of the so-called directional plasmons.   The obtained results  are used to analyze conditions needed for realization  of the HD regime in the optical experiments.

\section{Model and basic equations} 

The purpose of the current work is to  calculate  a contribution of thermoelectric   effects into the ratchet current and  also to take into account effects related to finite viscosity.

We will focus on the ratchet effect in 2D electron gas with a parabolic energy spectrum which is  covered by  a   long periodic (with a period $L$) grating gate 
%(GGS)
with an asymmetric configuration of gate electrodes, e.g.,
dual-grating top gate (DGG) structures~\cite{Olbrich2011,Otsuji2013,BoubangaTombet2014,Faltermeier2015,Olbrich2016}.    
Application of the voltages to the grating electrodes   leads to  the static  periodic  potential in the 2D gas,
\be
%\label{V}
U(x)   = U_0\cos (q x +\phi), 
\label{U}
\ee
which modulates the electron density in the channel. 
Here  $ q={2\pi}/{L}$ is the modulation  wave-vector.

The system is  illuminated by an external radiation with a large wavelength, $\lambda \gg L,$ which is linearly  polarized in  $x$  direction.  The grating leads to modulation of this field  with the depth $h.$ In particular, for linear polarization of the incoming radiation along the $x$ axis,   
 the field, $\mathbf E (x,t)= E(x,t)  \mathbf e_x$ ($\mathbf e_x$ is the unit vector in the $x$ direction), acting in the 2D channel,
has spatially modulated amplitude:
\be
 %\label{Ex}
 E(x,t)  =  [1+ h \cos(q x)]  E_0   \cos \omega t .
\label{E}
\ee
We assume below that $h\ll 1.$ Following Ref.~\cite{Ivchenko2011}, we take into account  the asymmetry of the structure phenomenologically, by introducing the phase shift   $\phi$ between the static potential and radiation modulations.

The response of the  2D electron system to the above described perturbation
can be found by using two approaches -- hydrodynamic 
%(HD)  
and drift-diffusion. 
%(DD).
In both approaches, the direction of the radiation-induced dc  current is controlled by the lateral asymmetry parameter~\cite{Ivchenko2011}
%The ratchet current is governed by the following average over the ratchet period
%
\begin{equation}
\label{Xi}
\Xi = \left \langle |\mathbf  E(x,t)|^2 {\mathrm{d}U(x)\over \mathrm{d}x}\right \rangle_{t,x} = \frac{E_0^2 h U_0 q \sin \phi  }{2} ,
\end{equation}
where  averaging is taken  over the   period $2\pi/\omega $ and distance  $L.$

 Actually, the effect of the ee-interaction is twofold and can be quite strong. First of all, as we mentioned, suciently fast ee-collisions can drive the system into the hydrodynamic regime. Secondly, ee-interaction leads to plasmonic oscillations, so that a new frequency scale, the plasma frequency, $\omega_p (q)$ appears in the problem. The ratchet effect is dramatically enhanced in the vicinity of plasmonic resonances. 
%%%%
%%%%
%\subsection{Basic %equations}
%%%%
%%%%
We  start with a system of hydrodynamic equations, 
\begin{align}
     & \frac{\partial N}{\partial t}+\frac{\partial}{\partial x} \left(N v\right)=0, 
     \label{N}
     \\ 
     & \frac{\partial v}{\partial t}+v \frac{\p v}{ \p x} +\frac{v}{\tau}-\eta\frac{\p^2 v}{\p^2 x} ={a}-s_0^2 \frac{\p n}{\p x}  - \frac{1}{m N}\frac{\p W}{\p x} , 
     \label{v}
     \\ 
     &  \mathcal{C}\left[\frac{\partial T}{\partial t}+\frac{(T {v})}{\p x}\right]=N \left(\frac{T_0-T}{\tau_{ph}}+\frac{m v^2}{\tau}\right),
     \label{T}
\end{align}
describing evolution  of three local variables: 
concentration, $N= N(x,t),$  electron fluid velocity $ v =  v (x,t),$ and  temperature $T=T(x, t).$ 
Here $T_0$ is the lattice temperature,
$$a=\frac{e}{m}\left(E-\frac{ \p U}{ \p x}\right)$$
is the total force acting on electron from  both the static potential and radiation field,
$s_0= \sqrt{e U_{\rm g}/m}$ is the plasma wave velocity, which is controlled by  the backgate voltage $U_{\rm g},$     $n=(N-N_{0})/N_0$ is the dimensionless concentration,   $N_0$ is the equilibrium electron concentration   in 2D channel, $\eta$ is the electron viscosity, and $W$  is the  local energy of the Fermi gas in the moving frame, 
\begin{equation}
   W=\nu \int  \varepsilon \left[e^{(\varepsilon-\mu)/T+1}\right]^{-1}d\varepsilon\approx\frac{N^2}{2\nu}+\frac{\nu T^2 \pi^2}{6},
\label{W}
\end{equation}
where we took into account that for  2D gas the local concentration is connected with  local chemical potential, $\mu,$ as follows: $N= \nu \mu.$ Equation  \eqref{v} contains also  viscous friction  $\eta \p^2 v/\p^2x$ (we neglected here a concentration dependence of this term), where $\eta$ is viscosity.

Calculating the spatial derivative of the first term in $W,$  substituting it into Eq.~\eqref{v} and combining with the term $s_0^2 \p n/\p x $,  we find correction to  the plasma wave velocity due to pressure of the electron liquid:
\be
s_0 \to s=\sqrt{ s_0^2  + v_{\rm F}^2/2,}
\label{vF-in-s}
\ee
where $v_{\rm F}$ is the Fermi velocity.  
Next, calculating spatial derivative of the second  term in $W$  and  substituting it into Eq.~\eqref{v}
 we find the thermoelectric correction to the r.h.s. of Eq.~\eqref{v}
 \be
 a_{\rm th}= -\frac{\pi^2}{3 m } \frac{T}{\mu} \frac{\p T}{\p x}= -\frac{\mathcal C}{m N_0 (1+n)}\frac{\p T}{\p x}.
 \label{ath}
 \ee
Due to the term \eqref{ath},   Eq.~\eqref{T} couples with  Eqs.~\eqref{N} and \eqref{v}.  This coupling was neglected  in Refs.~\cite{Rozhansky2015,Moench2022}.

Following Refs.~\cite{Rozhansky2015,Moench2022}, one can solve Eqs.~\eqref{N}, \eqref{v}, and \eqref{T} perturbatively with respect to $E_0$ and $U_0.$  Non-zero radiation-induced dc current    appears  in the third  order (second order with respect to the radiation field and the first order with respect to static potential): $j_{\rm dc} \propto \Xi,$
where $\Xi$ is given by Eq.~\eqref{Xi}.  

\section{Calculation}
Below, we calculate separately the thermoelectric  and  viscous contributions.
%%%%%%%
%%%%%%
\subsection{Thermoelectric contribution}
%%%%%%%
%%%%%
\subsubsection{Linear polarization}
In this section, we put  $\eta=0 $ and   discuss a  simplified approach to solution of  Eqs.~\eqref{N}--\eqref{T}. We will start with a discussion of the linear polarization and then generalize the result for the case of arbitrary polarization.  
%\subsubsection{Linear polarization %along $x$ axis}
  First, we notice that  l.h.s. and r.h.s. of Eq.~\eqref{T} are  proportional  to  $\mathcal C \propto T, $  and $N\propto \mu,$ respectively. Hence, one can expect that for strongly degenerated Fermi gas, $T/\mu \ll 1,$  l.h.s. of Eq.~\eqref{T} can be neglected, so that temperature is fully determined by the balance between  local Joule heat and phonon  cooling
\be 
\frac{T-T_0}{\tau_{\rm ph}}=\frac{m v^2}{\tau}.
\ee
Expressing temperature from  this equation and substituting into  Eq.~\eqref{v}, we get a  system  of closed  equations for $n$ and $v$:  
\begin{align}
 \label{nnn}
    & \frac{\partial n}{\partial t}
    + 
    \frac{\partial [(1+n )v]}{\partial x}=0,
    \\
    \label{vvv}
    & \frac{\partial v}{\partial t}+v  \frac{\partial  v}{\partial {x}} \left[1+\frac{2}{\Gamma (1+n)}\right] +\frac{v}{\tau}=a -s^2 \frac{\partial n}{ \partial x } .
\end{align}
Here and in what follows we approximate  $\Gamma$ by its value at $\varepsilon=\mu.$        
We will search for  perturbative solution of Eqs.~\eqref{nnn} and \eqref{vvv} up to the third order with respect to $a$: 
$$ n=n^{(0,1)}+n^{(1,0)}+\cdots, \quad v=v^{(0,1)}+{v}^{(1,0)}+\cdots,$$ 
where the first and second indices  denote  an order of expansion with respect to $E_0$ and $U_0.$  This expansion aims to calculate the averaged dc current $J_{\rm rat}= ({e \tau N_0}/{m})\left\langle (1+n) a  \right\rangle_{t,x}.$ 
%where average %over $t$ and %$x$ is taken %over time  %period  $2\pi/%\omega$ and  %distance $L.$ 
A non-zero response appears in the order $(2,1)$.    The calculation can be simplified to the purturbative expansion of  the second order by using a simple identity, which can be obtained by multiplying  Eq.~\eqref{nnn}  with $v,$  \eqref{vvv} with $(1+n),$ summing thus obtained equations and averaging over $x$ and $t.$ Doing so, we get      
\begin{equation}
\label{Jratn}
 J_{\rm rat}=  \frac{e \tau}{m}N_0 \left\langle n^{(1,1)}   E - n^{(2,0)} \frac{\partial U}{\partial x}  \right\rangle_{t,x},
\end{equation}
where  $U$  and $E$ are given by Eqs.~\eqref{U} and \eqref{E}, respectively. Hence, we  only need to calculate the concentration    up to the second order with respect to $a.$ 

The calculations are fully  analogous to the ones performed in  Refs~\cite{Rozhansky2015} and \cite{Moench2022}, so   that we delegate them into Appendix \ref{App:simplified}. The result reads
% \red{\begin{equation}
%     J_{dc,x}=\frac{e^3  \tau N_0 q h |E_0|^2 U_0 \sin{\phi}(\omega^4-\omega^2\omega_q^2+\gamma^2(\omega^2+\Gamma\omega_q^2))}{2 m^3 s^2 \Gamma(\gamma^2+\omega^2)(\omega^2\gamma^2+(\omega^2-\omega_q^2)^2)}
% \end{equation}}
% \red{!!!!Комментарий: предлагаю переписатть формулу 18 и 20 через размерные едиинцы. !!!!!}
\begin{equation}
    \frac{J_{{\rm rat},x}}{J_0}=\frac{\Omega^2(1+\Omega^2)+\Omega_q^2(\Gamma-\Omega^2)}{\Gamma(1+\Omega^2)(\Omega^2+(\Omega^2-\Omega_q^2)^2)},
\label{Jx-lin}
\end{equation}
where  
\be 
J_0 = \frac{e^3 \tau^3 N_0 q h |E_0|^2 U_0 \sin \phi}{4m^3 s^2}
\label{J0}
\ee 
%%%
and  we used  dimensionless  variables  $$\Omega=\omega\tau, \quad \Omega_q=\omega_q\tau. $$ 
For $\Gamma \to \infty$ we reproduce result of Ref.~\cite{Rozhansky2015}. One can easily combine this equation with formula (33) of Ref.~\cite{Moench2022}, where  thermoelectric effects were neglected (by assuming $\Gamma \gg 1$), while the ratchet current was calculated for an arbitrary relation between  $\tau_{ee}$ and $\tau.$ This yields: 
% \red{\begin{equation}
%      J_{dc,x}=\frac{e^3 \tau N_0  q h |E_0|^2 U_0 \sin{\phi}\left(\gamma^2\omega_q^2+\omega^2(\gamma^2+\omega^2-\omega_q^2)\left[\frac{1}{\Gamma}+\frac{\tau_{ee}}{2(\tau+\tau_{ee})}\right]\right)}{2 s^2 m^3 (\gamma^2+\omega^2)(\omega^2\gamma^2+(\omega^2-\omega_q^2)^2)}
% \end{equation}}
\begin{equation}
\label{j-rat-t*}
    \frac{J_{\rm{rat},x}}{J_0}=\frac{ \Omega_q^2 + \Omega^2 \left(1+\Omega^2- \Omega_q^2 \right) \left[ \frac{1}{ \Gamma} + \frac{\tau_{ee}}{2 (\tau+ \tau_{ee})}  \right]}{(1+\Omega^2)(\Omega^2+(\Omega^2-\Omega_q^2)^2)}.
\end{equation}
%\Omega^2(1+\Omega^2)+\Omega_q^2(\Gamma-\Omega^2)
Expression in the square brackets can be written as
\be
  \frac{1}{ \Gamma} + \frac{\tau_{ee}}{2 (\tau+ \tau_{ee})}
 = \frac{\tau_{\rm T} + \tau_*}{ \tau}, 
  \ee
where $1/(2 \tau_*) =1/\tau+ 1/\tau_{ee}.$ As seen, there are two terms in Eq.~\eqref{j-rat-t*} with different asymptotic  behavior at $\Omega \to  \infty$: first, containing $\Omega_q^2$ in the numerator, scales as $1/\Omega^6$,  and  the second,  proportional     to the square bracket in the numerator with $1/\Omega^2$  high-frequency scaling.   Physically, the second term  is related  with the energy relaxation processes---temperature relaxation  and collision-induced thermalization.

\subsubsection{Arbitrary polarization}

The above result can be easily generalized for the case 
  of an arbitrary polarization  of the external field  described by phases $\alpha$ and $\theta$:
  \begin{align}
\label{Ex}
E_x (x,t) & = [1+ h \cos(q x+\phi)]  E_0 \cos \alpha  \cos \omega t ,
\\
\label{Ey}
E_{y}(x,t) & = [1+ h \cos(q x+\phi)]  E_0 \sin \alpha  \cos (\omega t+\theta).
\end{align}
% \be
% E_x = {E_{0}} \cos \alpha \cos {\omega}t, ~~
% E_y = E_{0}\sin \alpha \cos \left( {{\omega}t + \theta } \right).
% \label{Eext}
% \ee
These phases are related to the standard Stokes parameters
as follows:
$ P_{\rm L1}=\cos{2\alpha},$  $P_{\rm L2}=\sin{2\alpha} \, \cos \theta, $ $
		P_{\rm C}=\sin{2\alpha} \, \sin \theta. $
% \be
% \begin{aligned}
% 	& P_{\rm L1}=\cos{2\alpha},
% \\
% 	& P_{\rm L2}=\sin{2\alpha} \, \cos \theta,\\
% 		&P_{\rm C}=\sin{2\alpha} \, \sin \theta.
% \label{Stockes}
% \end{aligned}
% % \ee
% The components of total field in the channel read
 Calculations  of the current for this case are presented in  Applendix~\ref{App:arbitrary}. The result for the $x$ component of the current  modifies  as follows
 %, namely, the factor 
% $ cos^2 \alpha = (1+P_{\rm L1})/2 $ appear. 
% \begin{equation}
%     \label{j-ratx}
%  \frac{J_{{\rm rat},x}}{J_{0}}=\frac{\Omega^2(1+\Omega^2)+\Omega_q^2(\Gamma-\Omega^2)}{ \Gamma(1+\Omega^2)(\Omega^2+(\Omega^2-\Omega_q^2)^2)}(1+P_{L1})
% \end{equation}
% \red{NEW!!!\begin{equation}
%     \label{j-ratx}
%  \frac{J_{{\rm rat},x}}{J_{0}}=\frac{P_0[2\Omega^2(1+\Omega^2)+\Omega_q^2(\Gamma-3\Omega^2)+\Omega_q^4]+P_{L1}\Omega_q^2(\Gamma+\Omega^2-\Omega_q^2)}{ \Gamma(1+\Omega^2)(\Omega^2+(\Omega^2-\Omega_q^2)^2)}
% \end{equation}}
% % \begin{align}
% % &\label{j-ratx}
% %  J_{{\rm rat},x}=\frac{e^3 \tau N_0  q h|E_0|^2 U_0 \sin{\phi}(\omega^4-\omega^2\omega_{q}^2+\gamma^2(\omega^2+\Gamma\omega_q^2))}{4 m^3 s^2 \Gamma(\gamma^2+\omega^2)(\gamma^2\omega^2+(\omega^2-\omega_q^2)^2)}(1+P_{L1})
% % \end{align}
% %%%%%%%%
% %%%%%%%%%%
 % $$\frac{\text{$\Omega $q}^2 \cos ^2(\alpha ) \left(\frac{\Omega ^2-\text{$\Omega $q}^2}{\Gamma }+1\right)}{\left(\Omega ^2+1\right) \left(\left(\Omega ^2-\text{$\Omega $q}^2\right)^2+\Omega ^2\right)}+\frac{1}{\Gamma  \left(\Omega ^2+1\right)}$$

\be
\begin{aligned}
\frac{J_{{\rm rat},x}}{J_{0}}&= \frac{1}{\Gamma  \left(\Omega ^2+1\right)}
\\
&+\frac{\Omega_q^2 \left[1+(\Omega ^2-\Omega_q^2)/\Gamma \right]   \cos^2\alpha }{\left(\Omega ^2+1\right) \left[\left(\Omega ^2-\Omega_q^2\right)^2+\Omega^2\right]}.
\end{aligned}
\label{j-ratx}
\ee
 For $\alpha=0,$ we restore Eq.~\eqref{Jx-lin}. 

There is also  $y$ component of the current 
\be
  \frac{J_{{\rm rat},y}}{J_0}
 =\frac{(\Omega_q^2-\Omega^2)P_{\rm L2}-\Omega P_{\rm C}}{2 (\Omega^2+(\Omega^2-\Omega_q^2)^2)}
\label{j-raty}
\ee
which is insensitive to the temperature relaxation and thermalization, and coincides with previously obtained result \cite{Rozhansky2015}.  
% \be
%   J_{{\rm rat},y}
%  =\frac{e^3 \tau N_0  q h|E_0|^2 U_0 \sin{\phi}((\omega_q^2-\omega^2)P_{L2}-\gamma\omega P_{c})}{8 m^3 s^2 (\gamma^2\omega^2+(\omega^2-\omega_q^2)^2)}
% \label{j-raty}
% \ee
%Here,   $J_0=(e^3\tau^3 N_0 q h |E_0|^2  U_0 \sin \phi)/(4 m^3 s^2)$

%Let us note two  features of Eqs.~\eqref{j-ratx} and \eqref{j-raty}. First, 

%\color{blue} !добавить  в кусок, написанный мажентой, ссылок и сослаться аккуратнее (Лёня)!
%\color{magenta}
We notice that 
  for $\Gamma=\infty $ we reproduce results of Ref.~\cite{Rozhansky2015}. 
 We also see that  
  in the absence of plasmonic effects, i.e. for 
 $\Omega_q \to 0$,  we get for the $x$ component of the current the polarization-independent (Seebeck) contribution to the ratchet effect in the non-interaction system  \cite{Ivchenko2011, Nalitov2012, Budkin2016a}:  
\begin{equation}
    \label{j-ratx-seebeck}
 \frac{J_{{\rm rat},x}}{J_{0}}\biggr|_{\rm non-int}= \frac{ 1}{\Gamma (1+\Omega^2)}.
\end{equation}
It is worth noting that in the DD approximation  when  $\tau_{\rm ee} = \infty$, and for $\Omega_q=0$ there is also a polarization-dependent contribution to the ratchet current, so that the $x$ component of the total current within DD approach  reads~\cite{Moench2023} 
\begin{equation}
    \label{j-DD-ratx-seebeck}
 \frac{J^{\rm DD}_{{\rm rat},x}}{J_{0}}= \frac{ 1}{ 1+\Omega^2}\left(  \frac{1}{\Gamma} + \frac{\cos^2\alpha}{2}\right).  
\end{equation}
%First term  was obtained in Ref.~\cite{},   while second in Ref.~\cite{}.
At  $\alpha=0$ this equation agrees with    Eq.~\eqref{j-rat-t*} taken for ${\tau_{\rm ee} \to \infty}$,~$\Omega_q=0$.   Two important comments are needed here: (i) Since we compare our results  with the DD approach,  we need to ``switch off'' the interaction.  This means  that  we have to replace $s^2$ with  $v_{\rm F}^2/2$ [see Eq.~\eqref{vF-in-s}]  in  the definition of $J_0$ [see Eq.~\eqref{J0}]; (ii)~the frequency $\Omega_q$ is neglected 
%Frequency $\Omega_q$ does not turns to zero exactly after above mentioned  replacement. However,  since typically 
%$v_{\rm F} q$ is sufficiently   small ($v_{\rm F} q \ll \omega,~v_{\rm F} q \ll \gamma$)  
%$s \gg v_{\rm F},$ 
%one can neglect in the DD regime $\Omega_q $ with respect to $\Omega$ 
both in the polarization-dependent and polarization-independent response.  
Having these comments in mind, we conclude that  our theory  includes DD approximation  as a limiting case.  
%\color{black}

Analyzing the above derived equations we  conclude  that  the thermoelectric contribution calculated with account  for the  plasmonic effects dramatically changes the   frequency  dependence of the response.   In the next subsection we analyze this dependence in more detail.

\subsubsection{Illustration of different regimes}

Thermoelectric effects do not change $J_{{\rm rat},y},$ so that  its dependence on $\Omega$ is the same as was discussed in Ref.~\cite{Rozhansky2015}. However, the frequency behavior of the $x$  component of the current dramatically changes with decreasing  $\Gamma$ and can change sign in some interval of $\Omega.$  In particular,  this can be seen  in the resonant approximation.  By replacing  in  Eq.~\eqref{j-rat-t*}, 
$\Omega = \Omega_q + \delta \Omega,$   and assuming $\Omega_q$ very large for fixed $\delta \Omega = (\omega-\omega_q) \tau, $ we get
\be
J_{\rm{rat},x} \approx  \frac{J_0}{\Omega_q^2}    \frac{1+  2 \delta \Omega  ~\omega_q (\tau_{\rm T} + \tau_*)  }{1+ 4 \delta \Omega^2}. 
\label{Jrat-asymm}
\ee
As seen, a new parameter, $ \omega_q (\tau_{\rm T} + \tau_*),$ appears in the problem. For not too large $\Gamma$ this parameter becomes large,  so that current becomes odd function of $\delta \omega,$ which is negative approximately in all region $\delta \Omega <0.$       
 \begin{figure}[h!]
	\centering
	\includegraphics[angle=0,width=1.1\linewidth]{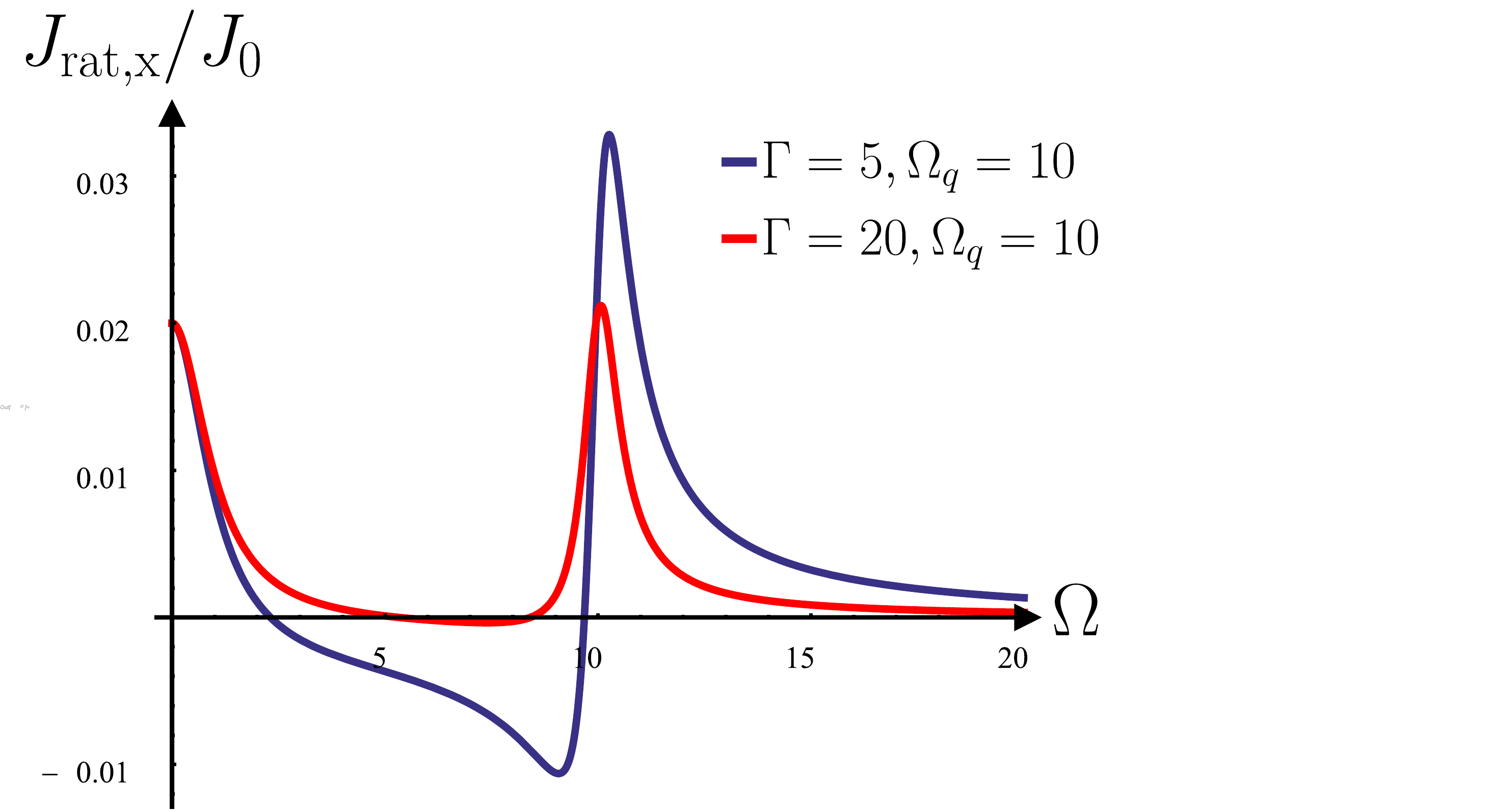}
	\caption{Dependence of the ratchet current on dimensionless frequency, $\Omega=\omega \tau,$  at  two values of  $\Gamma.$  For  $\Gamma=20$ (red curve) there are two peaks: the Drude peak at $\Omega=0$ and the plasmonic resonance at $\Omega=\Omega_q$  in agreement with Ref.~\cite{Rozhansky2015}.  With decreasing  $\Gamma$ up to lower value $\Gamma=5$ (blue curve),  the plasmonic resonance  modifies and becomes strongly asymmetric [see Eq.~\eqref{Jrat-asymm}].    }
	\label{Fig1}
\end{figure}
 \begin{figure}[h!]
	\centering
	\includegraphics[angle=0,width=1.1\linewidth]{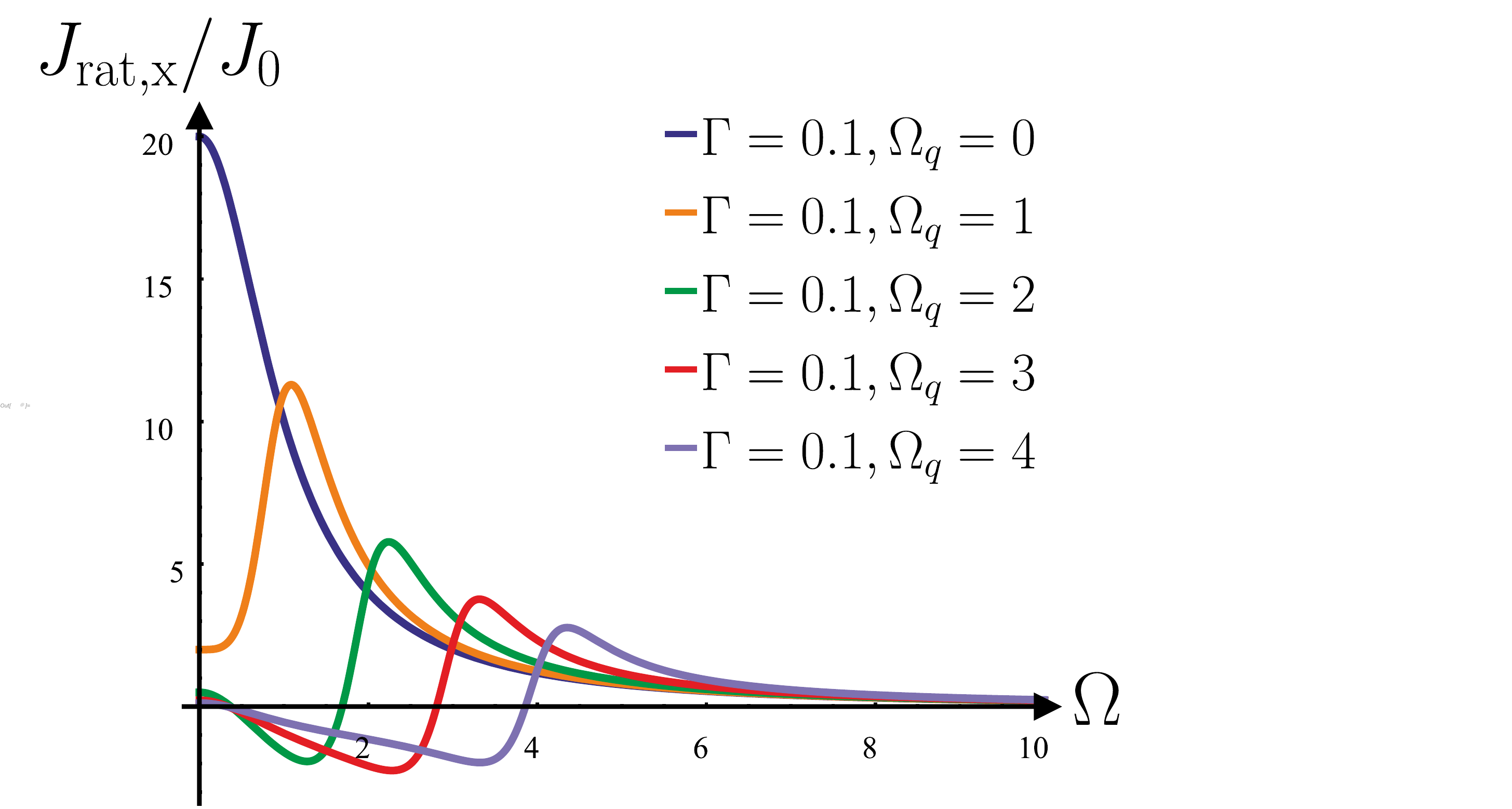}
	\caption{Dependence of the ratchet current   on the  dimensionless frequency for fixed small $\Gamma$ and different values of the quality factor of the plasmonic resonance, $\Omega_q.$ For $\Omega_q=0,$ plasmonic resonance is absent and response shows the Drude resonance at $\Omega=0.$  With increasing the quality factor, the  asymmetric plasmonic peak appears.    }
	\label{Fig2}
\end{figure}
 \begin{figure}[h!]
	\centering
	\includegraphics[angle=0,width=1.1\linewidth]{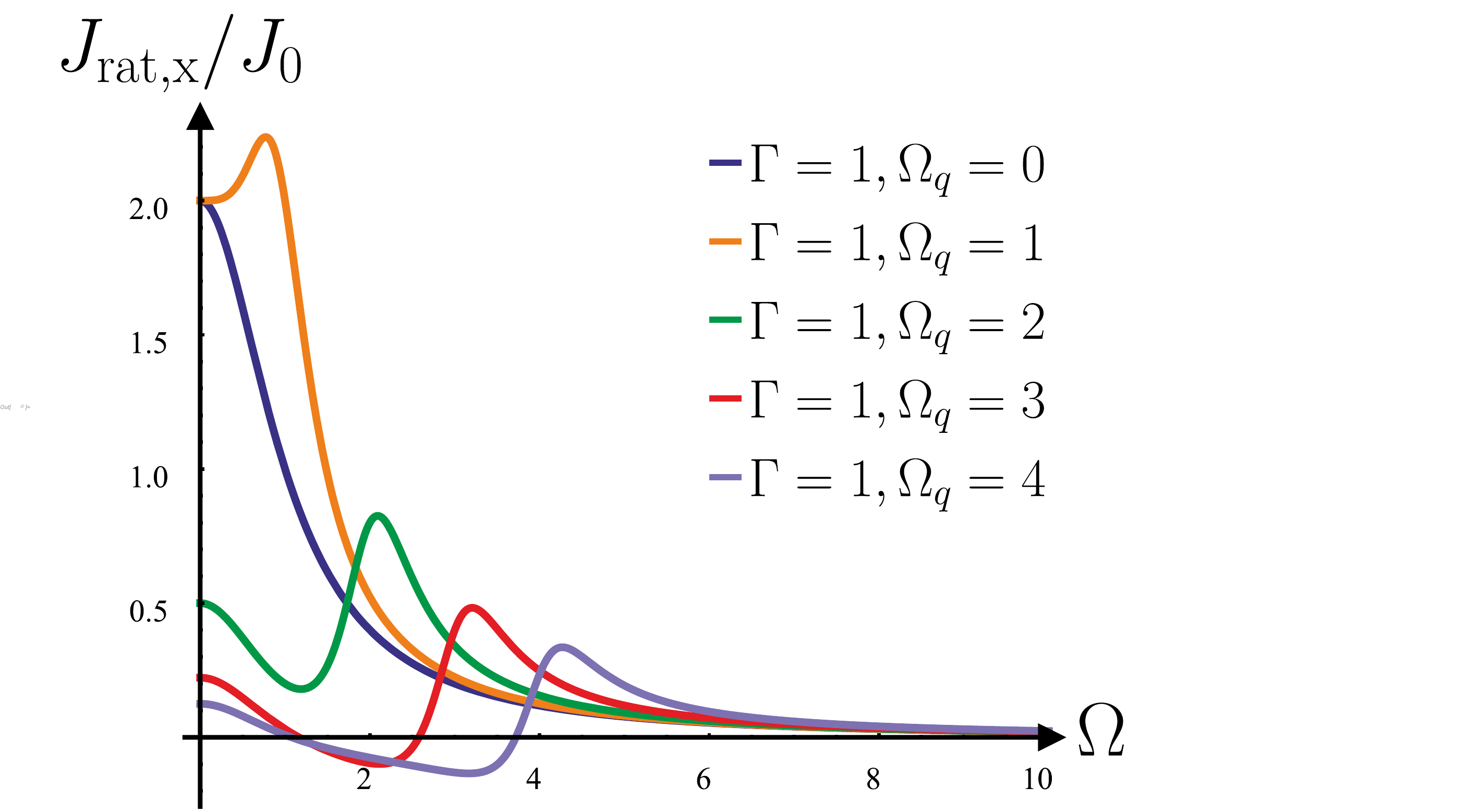}
	\caption{ The same as in Fig.~\ref{Fig2} but at lrger value of $\Gamma.$            }
	\label{Fig3}
\end{figure}
 \begin{figure}[h!]
	\centering
	\includegraphics[angle=0,width=1.1\linewidth]{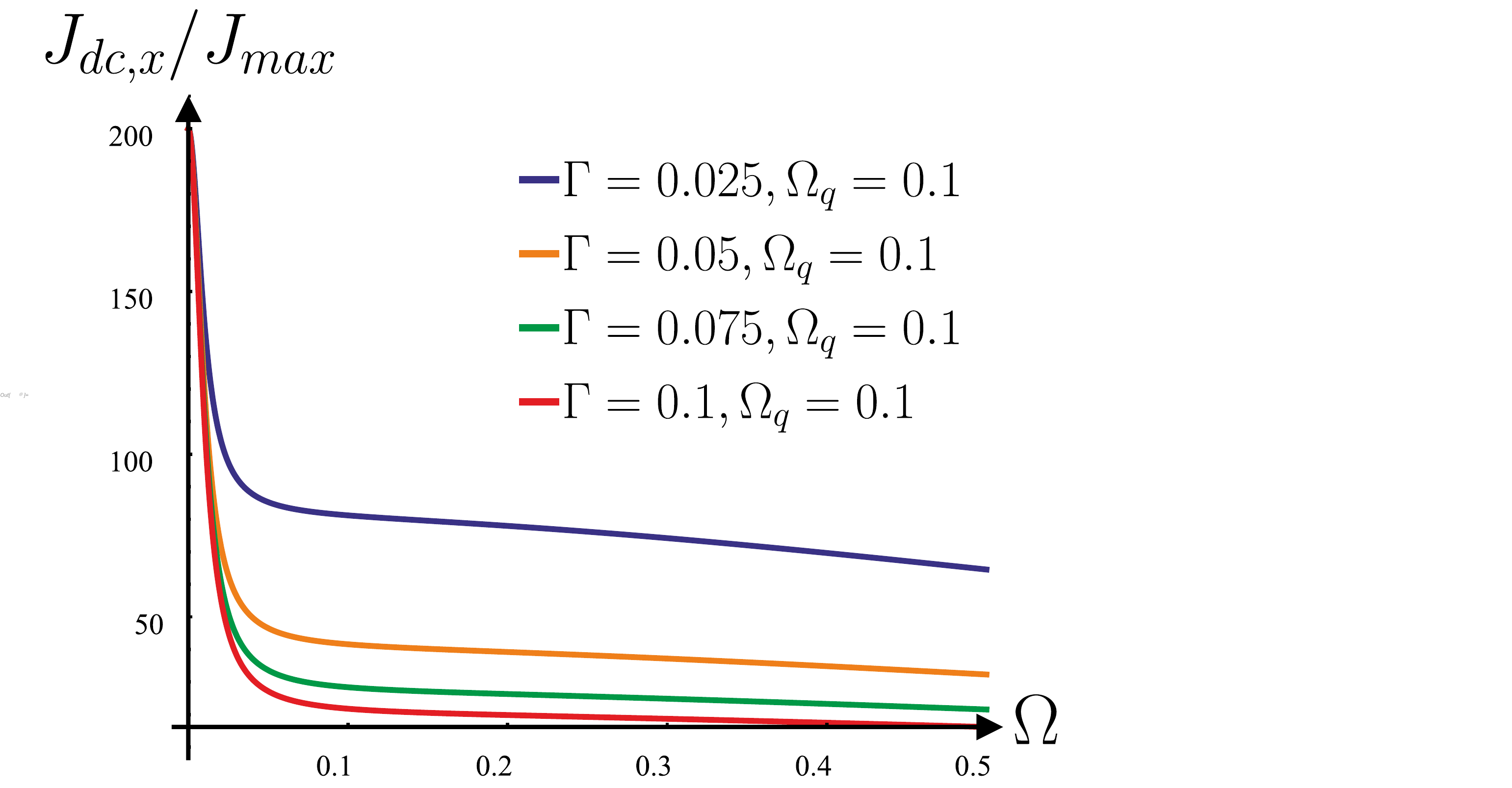}
	\caption{Non-resonant response for low quality factor $\Omega_q=0.1$ and small values of $\Gamma.$  Narrow peak   at $\Omega=0$  
 arises on the top of the smooth Drude peak 
  due to the Maxwell relaxation      }
	\label{Fig4}
\end{figure}

Dependence of the ratchet current on the dimensionless  frequency for various values of  $\Gamma$ and $\Omega_q$ is shown in Figs.~\ref{Fig1},~\ref{Fig2},~\ref{Fig3} for the case of linear polarization along the $x$ axis.

In Fig.~\ref{Fig1} we plot  the ratchet current as a function of the dimensionless frequency $\Omega$  for  two values of  $\Gamma.$  For large value  $\Gamma=20$  the results of Ref.~\cite{Rozhansky2015}  are reproduced. Specifically,   dc response shows  the Drude peak at $\Omega=0$  and the  plasmonic resonance at $\Omega=\Omega_q$  in agreement with Ref.~\cite{Rozhansky2015}.  With decreasing  $\Gamma$   the plasmonic resonance  becomes asymmetric  and changes sign in some interval  of the frequencies. The evolution of the resonance from symmetric to asymmetric shape is well described by    Eq.~\eqref{Jrat-asymm}.

Figures~\ref{Fig2} and \ref{Fig3} illustrate the  plasmonic effect in the regime of small $\Gamma,$ when thermoelectric correction dominates.  At zero plasmonic frequency, $\Omega_q=0,$ the dc response is given by  the polarization-independent Seebeck peak having the Drude shape [see, Eq.~\eqref{j-ratx-seebeck}].  For sufficiently  large  quality factor, $\Omega_q \geq 1,$ there appear an asymmetric plasmonic resonance at $\Omega =\Omega_q.$  

Most interesting behavior is obtained  at small $\Gamma,$ when thermoelectric contribution is large,   in the non-resonant regime, for $\Omega_q \ll 1.$ This case is illustrated in Fig.~\ref{Fig4}. As one can see a very narrow peak appears on the top of the smooth Drude dependence.  Physics behind this peak is the Maxwell charge relaxation, which has a characteristic frequency scale $\Omega \sim \Omega_q^2 \ll 1,$ i.e.  for $\omega \sim \omega_q^2/\gamma \ll \gamma$  in dimensional units. For 
$$ \Omega_q \ll       \sqrt{\Gamma} \ll 1,$$  analytical expression   describing Fig.~\ref{Fig4} can be found from Eq.~\eqref{j-rat-t*} (where we put for simplicity $\tau_{\rm ee}=0$
):   
\begin{equation}
\frac{J_{\rm{rat},x}} {J_0} \approx \frac{\Omega_q^2}{\Omega^2+ \Omega_q^4}+
    \frac{1}{\Gamma} 
      \frac{1}{1+\Omega^2}. 
\label{Maxwell-rel}
\end{equation}
Two terms here represent   the  Maxwell relaxation peak \cite{Rozhansky2015} and  the Seebeck polarization-independent contribution

\section{Viscous contribution}
In this section we will discuss effect of finite viscosity of the electron liquid.  We limit ourselves with the case of the linear polarization along $x$ direction and assume $\tau_{\rm ee} \ll \tau.$    Then, the electron liquid velocity is described by the Navier-Stokes equation:   
 \be
    % & \frac{\partial n}{\partial t}+\nabla \mathbf{v}=-\nabla \left\left(n \mathbf{v}\right),
    %  \label{nneta}
    %  \\
    % & 
    \frac{\partial {v}}{\partial t} +\frac{{v}}{\tau}+s^2 \frac{ \partial n}{\partial x}-\eta\frac{\p^2 v}{\p^2 x}=a-v  \frac{\p v}{\p x}\left(1+\frac{2}{\Gamma}\right),
    \label{vveta}
\ee
We demonstrate below that effect of   viscosity on the electronic ratchet effect  is small and give only small correction to the $J_{{\rm dc},x}.$   However, the  viscosity gives the key  contribution to the excitation of the  so-called  travelling directional plasmons, which can be excited along with the ratchet dc current  provided grating gate structure does not have inversion center   \cite{Popov2015,
Fateev2019,
Moiseenko2020,
Morozov2021}.

\subsection{Small correction to the electronic ratchet current}
Next, we discuss viscosity-induced correction to the electronic ratchet.
We take into account that viscosity has dispersion described by the following equation \cite{Alekseev2018}:
\be
\eta \to \eta(\omega)= \frac{\eta(0)} {1-i \omega \tau_{\rm ee}}.
\label{eta-omega}
\ee

The calculations are fully analogous to  the ones presented in  Appendix \ref{App-lin}.   The  details of calculations  of the viscosity-induced correction as well as the most general formula for the dc ratchet current  with account of the  viscosity, Eq.~\eqref{J-eta},   are  presented in Appendix~\ref{App-vis}.  Analyzing Eq.~\eqref{J-eta} one can see that viscosity gives a small contribution to the ratchet current in the whole frequency interval. Here, we illustrate   it for the case of high-frequency asymptotic. To this end we expand  Eq.~\eqref{J-eta} up to linear order with respect to $\eta$ and assume that $\Omega \gg \Omega_q.$ The linear in $\eta$ term contains factor $Q^2 \ll 1.$ We neglect $Q$ in this term everywhere except this factor. Then, we obtain      
\be
\begin{aligned}\frac{J_{\rm{rat},x}} {J_0} &\approx \frac{\Omega_q^2}{\Omega ^6}+ \frac{1}{\Omega^2} \left( \frac{1}{\Gamma} + \frac{\tau_{\rm ee}}{2}  \right) 
\\
& + \frac{\eta Q^2}{1+\Omega^2 \tau_{\rm ee}^2} \left ( \frac{1- \tau_{\rm ee} \Omega^2}{\Gamma \Omega^4} - \frac{\tau_{\rm ee}}{2\Omega^2}\right),
\end{aligned}
\ee
where  we  use dimensionless units for $\tau_{\rm ee}$ (measured in the units of  $\tau$) and $\eta$ (measured in the units of $v_{\rm F}^2 \tau.$).

The first line of this equation represents  asymptotic  of  Eq.~\eqref{j-rat-t*} at $\Omega \to \infty. $  The second line represents  viscosity-induced correction. The collision time $\tau_{\rm ee}$ appears in  this correction  due to dispersion of $\eta.$   It is worth noting that at $\Gamma \to \infty$ viscosity term in non-zero only due to this dispersion.  Having in mind that dimensionless $\eta$ equals to $\tau_{ee}/4\tau,$ one can easily see that  viscosity gives a small correction for any $\Gamma.$   At very high frequency this correction scales as $1/\Omega^4.$

\subsection{Viscosity-driven directional plasmons }
In this subsection, we demonstrate that viscosity of the electron liquid, alhtough yielding a small correction to the electron ratchet effect, fully determines excitation of travelling directional plasmons \cite{Popov2015,
Fateev2019,
Moiseenko2020,
Morozov2021}.  

Physics behind directional  plasmons is as follows. Let us consider  plasmonic oscillations of the density in the channel.    Perturbation theory yields  two terms which oscillate both in $t$  and in $x$: term of the first order  $n^{(1,0)}$ and  the second order term $n^{(1,1)}.$  The sum of these terms can be presented as two waves propagating in the opposite directions       
\begin{equation}
  \begin{aligned}
    n^{(1,0)}+n^{(1,1)}=&C_1 \cos (q x-\omega t-\alpha_1)
    \\
    &+C_2 \cos(q x+\omega t-\alpha_2),
\end{aligned}
\end{equation}
where $C_{1,2} $ and $\alpha_{1,2}$ are the amplitudes and phases of these waves, respectively.  

Since the grating gate structure does  not have inversion center, the symmetry of the problem allows for  $C_1 \neq C_2.$  Then, the number of right- and left-moving plasmons is different and one can  define  
\begin{equation}
\label{Jpl}
    S_{\rm{pl}}=s(C_1^2-C_2^2)
\end{equation}
as a flux of directional  plasmons travelling in a certain direction.   
 Calculation of $S_{\rm pl}$ is presented in  the Appendix \ref{App: directed pl}.   In the limit $\eta \to 0,$  the   result reads  
\begin{equation}
\frac{S_{\rm{pl}}}{S_0}=\frac{\Omega}{(1+\Omega^2)((\Omega^2-\Omega_q^2)^2+\Omega^2)}, \label{Spl}
\end{equation}
%%%%
where \be S_{0}=\frac{e^3 \tau^5 q^4 h |E_0|^2 U_0 \eta \sin\phi}{m^3 s} \propto \eta~ \Xi \ee
We see that  $  S_{0}$ is proportional to $\eta$ (we neglect here terms of the order of $\eta^2$ and higher order).  Hence, remarkably, the  directed  travelling plasmons are absent in the ideal electron liquid and  appear only due to viscosity. We also see that $S_0$ as expected is proportional to the asymmetry parameter $\Xi.$ We also notice that $ S_{\rm{pl}}$    does not depend on $\Gamma.$

 \begin{figure}[h!]
	\centering
	\includegraphics[angle=0,width=1.1\linewidth]{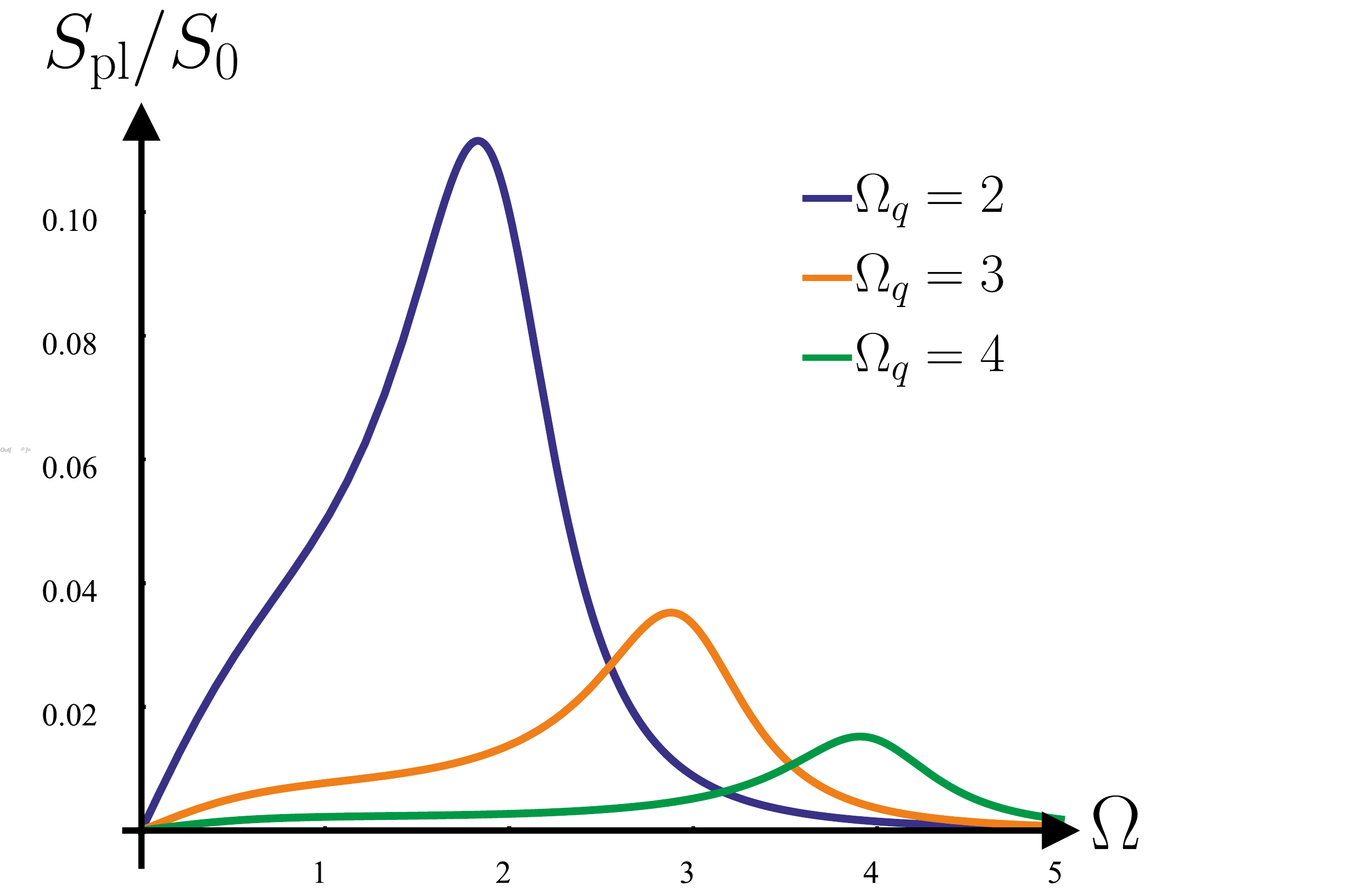}
	\caption{Dependence of the directed plasmons flux on the dimensionless frequency for different quality factors  }
	\label{Fig-Spl}
\end{figure}

Dependence of the directed plasmons flux on the dimensionless frequency is plotted in Fig.~\ref{Fig-Spl}. This dependence contains symmetric plasmonic resonsnce similar to the resonance is dc ratchet current for $\Gamma \to \infty$. However, in contrast to electronic ratchet, the resonance at $\Omega =0$  is absent.

\section{Conclusion}
To conclude, we have  studied  thermoelectric and viscous  contributions  to the radiation-induced      electron  ratchet current $J_{\rm rat}$ and to the flux of the travelling directional plasmons, $S_{\rm pl},$  in asymmetric  dual-grating gate structure  without inversion center.  We  have demonstrated that thermoelectric effects  do not change  $S_{\rm pl}$  but dramatically modify $J_{\rm rat}.$ By contrast, viscosity of the electron liquid leads to a small  correction to    $J_{\rm rat}$  but fully determines the  flux of directed plasmons.

\section*{Acknowledgments}
We thank D.~V. ~Fateev and I.~V.~Gorbenko for useful discussions. The work  was supported by the 
Russian Science Foundation under grant 24-62-00010.
 S.O.P. also thank the Foundation for the Advancement of Theoretical Physics and Mathematics ``BASIS''.
 L.E.G. acknowledges the support of the Deutsche Forschungsgemeinschaft (DFG, German Research Foundation) via Project-ID 448955585 (Ga501/18).

$$$$

\appendix 
\begin{widetext}

 \section{ Calculation of the ratchet current within a simplified method for $T \ll E_{\rm F}$} \label{App:simplified}
 \subsection{Linear polarization}\label{App-lin}
 In this Appendix, we present derivation of photo-induced  dc current,  $J_{\rm rat},$ within a simplified approach, which is valid in the strongly degenerated Fermi gas,  when   $T \ll E_{\rm F}.$   The approach is based on perturbative solution of simplified HD equations  Eq.~\eqref{nnn} and Eq.~\eqref{vvv} with respect to external perturbation  $a$.  DC current appears in the order (2,1), i.e.  is proportional to $h E_0^2 U_0,$ and can be presented as a sum  two terms:
 \begin{equation}
  J_{{\rm rat}, x}=\frac{e \tau}{m}N_0\left\langle n a_x \right\rangle^{(2,1)}_{t,x}=J_{{\rm rat},x}^{\rm I}+J_{{\rm rat },x}^{\rm II},
  \label{Idc}
\end{equation}
where
\begin{align}
  & J_{{\rm rat},x}^{\rm I}=-\frac{e \tau}{m}N_0\left\langle n^{(2,0)} \frac{\partial U}{\partial x} \right\rangle_{t,x}, 
  \label{J-I}
  \\
  \label{J-II}
  & J_{{\rm rat},x}^{\rm II}=\frac{e \tau}{m}N_0 \left\langle n^{(1,1)} E_x \right\rangle_{t,x}
\end{align}
Hence, instead of calculation  of the third order perturbative  contribution to the current it is sufficient to calculate the second order correction to the dimensionless concentration  $n.$ 

We start with calculation of the linear response: $n^{(1,0)},n^{(0,1)},v^{(1,0)}, v^{(0,1)}$. Linearizing Eqs.~\eqref{nnn} and ~\eqref{vvv},  we find:
%response to  the linear perturbation %$a_x$:
\be
\label{1-order}
  \begin{aligned}
  &  n^{(1,0)}=-\frac{e h E_0 q\sin(q x)}{2m(\omega_q^2-\omega^2-i\gamma \omega)}e^{-i\omega t}+h.c.,
\\
   & v^{(1,0)}=-\frac{e E_0 }{2m(\gamma-i\omega)}e^{-i\omega t}-\frac{ e h E_0 \omega\cos(q x)}{2im(\omega_q^2-\omega^2-i\gamma \omega)}e^{-i\omega t}+h.c.,
\\
 & n^{(0,1)}=\frac{e V_0 }{m s^2}\cos(q x+ \theta),\\
&v^{(0,1)}=0.
\end{aligned}
\ee
  % \begin{equation}
 %     n_{\omega,q}^{(i,j)}=\frac{-i \hat{q}}{s^2\hat{q}^2-\hat{\omega}^2-i\hat{\omega}\gamma}\left[J_{v;q,\omega}^{(i,j)}-J_{n;q,\omega}^{(i,j)}(\gamma-i\hat{\omega})\right]
 % \end{equation}
 % \begin{equation}
 %     v_{\omega,q}^{(i,j)}=\frac{1}{s^2\hat{q}^2-\hat{\omega}^2-i\hat{\omega}\gamma}\left[J_{n;q,\omega}^{(i,j)}s^2\hat{q}^2+i \hat{\omega} J_{v;q,\omega}^{(i,j)}\right]
 % \end{equation}
%%%%%%%%%%%%
%%%%%%%
Substituting  Eq.~\eqref{1-order}  into nonlinear terms one can perform high order iteration   and do the next iterations  by using  the following matrix equation:
\be
 \begin{pmatrix}
    n \\  v 
\end{pmatrix}^{(i,j)}=\frac{1}{i s^2 \hat{q}^2+\hat{\omega}\gamma-i\hat{\omega}^2}\begin{pmatrix}
i\gamma-\hat{\omega} & \hat{q}\\ \hat{q} s^2 & \hat{\omega} 
\end{pmatrix}~\begin{pmatrix}
  J_n \\ J_v  
\end{pmatrix}^{(i,j)},  
 \ee
where   $\hat{q}= -i\partial_x$, $\hat{\omega}= -i\partial_t$ and  
\begin{equation}
\label{JvJn}
 \begin{pmatrix}
  J_n \\ J_v   
\end{pmatrix}^{(i,j)}=\begin{pmatrix}
   \partial_x (n v_x) \\ v_x \partial_x v_x\left(1+2/\Gamma\right) 
\end{pmatrix}^{(i,j)},   
\end{equation}
are non-linear sources arising  in the $(i,j)$ order of the perturbation expansion.  In order to calculate $n^{(2,0)}$ and $n^{(1,1)},$ we need to calculate   
\begin{equation}
    \left(\begin{array}{c} J_{n} \\ J_{v} \end{array}\right)^{(2,0)}=    \left(\begin{array}{c} -\partial_{x} (v^{(1,0)}n^{(1,0)}) \\  -v^{(1,0)} \partial_{x}v^{(1,0)}(1+2/\Gamma) \end{array}\right),
\end{equation}
and 
\begin{equation}
    \left(\begin{array}{c} J_{n} \\ J_{v} \end{array}\right)^{(1,1)}=  \left(\begin{array}{c} -\partial_{x} (v^{(1,0)}n^{(0,1)}) \\  0 \end{array}\right),
\end{equation}
where $n^{(1,0)},n^{(0,1)},v^{(1,0)}$ are given by Eq.~\eqref{1-order}. After some algebra, we get  
\be
\label{n20-n11}
\begin{aligned}
   &  n^{(2,0)}= n^{(2,0)}(x)=\frac{e^2 |E_0|^2 h \cos{(q x)}(-\gamma\Gamma+i(2+\Gamma)\omega)}{4 s^2 m^2 \Gamma(\gamma+i\omega)(\omega_q^2-\omega(i\gamma+\omega))}+h.c.,
 \\
    & n^{(1,1)}=n^{(1,1)}(x,t) =-\frac{e^2 E_0 U_0 q \sin{(q x+\theta)}}{2 s^2 m^2 (\omega_q^2-\omega(i\gamma+\omega))}e^{-i\omega t}+h.c..
 \end{aligned}
 \ee
Here, we  left only time-independent terms in $n^{(2,0)},$ skipping terms $\exp (\pm 2 i \omega t ),$ which do not give contribution to the $dc$ response. Substituting Eq.~\eqref{n20-n11} into Eqs.~\eqref{J-I} and \eqref{J-II}, we finally arrive at
 \begin{align}
   &  \frac{J^{\rm I}_{{\rm rat},x}}{J_0}=\frac{\left((2+\Gamma)\omega^2(\omega^2-\omega_q^2)+\gamma^2((2+\Gamma)\omega^2+\Gamma\omega_q^2)\right)}{\tau^2 \Gamma(\gamma^2+\omega^2)(\gamma^2\omega^2+(\omega^2-\omega_q^2)^2)},
 \\
    & \frac{J^{\rm II}_{{\rm rat},x}}{J_0}=\frac{\left(\omega_q^2-\omega^2\right)}{ \tau^2(\gamma^2\omega^2+(\omega^2-\omega_q^2)^2)},
 \end{align}
  \subsection{ Arbitrary polarization} \label{App:arbitrary}
  Calculations  presented  in  Appendix~\ref{App-lin} can be easily generalized for the case of arbitrary polarization, when the electric field is given by  Eq.~\eqref{Ex} and  Eq.~\eqref{Ey}. In this case, hydrodynamic equations involve also $y-$component  of the velocity: 
 \begin{align}
    & \frac{\partial n}{\partial t}+ \frac{\partial [(1+n)v_x]}{\partial  x} = 0,
     \label{np}
     \\
    &  \frac{\partial v_x }{\partial t} +\frac{v_x}{\tau}+s^2  \frac{\partial n}{\partial x}= a_x-\left(1+\frac{2}{\Gamma}\right) v_x \frac{\partial v_x}{\partial x}-\frac{2}{\Gamma}v_y\frac{\partial v_y}{\partial x},
    \label{vpx}
    \\
    &\frac{\partial v_y }{\partial t} +\frac{v_y}{\tau}= a_y- v_x \frac{\partial v_y}{ \partial x}  ,
    \label{vpy}
\end{align}
where $a_x=e/m (E_x-\partial U/\partial x)$,$a_y=e E_y/m $.  The calcaulations are fully analogous to  the ones presented in  Appendix \ref{App-lin}. 
Here, we present  most important modifications without going into technical details. We start with calculation of the linear response: $n^{(1,0)},n^{(0,1)},v_{x}^{(1,0)},v_{y}^{(1,0)}, v_{x}^{(0,1)}$. Linearizing Eqs.~\eqref{np}, ~\eqref{vpx} and~\eqref{vpy},  we find:
%response to  the linear perturbation %$a_x$:
\be
\label{1-order_Arb_pol}
  \begin{aligned}
  &  n^{(1,0)}=-\frac{e h E_{0x} q\sin(q x)}{2m(\omega_q^2-\omega^2-i\gamma \omega)}e^{-i\omega t}+h.c.,
\\
   & v_{x}^{(1,0)}=-\frac{e E_{0x} }{2m(\gamma-i\omega)}e^{-i\omega t}-\frac{ e h E_{0x} \omega\cos(q x)}{2im(\omega_q^2-\omega^2-i\gamma \omega)}e^{-i\omega t}+h.c.,
 \\
   & v_{y}^{(1,0)}=-\frac{e E_{0y} (1+h\cos(q x))}{2m(-\gamma+i\omega)}e^{-i(\omega t+\theta)}+h.c.,
\\
 & n^{(0,1)}=\frac{e V_0 }{m s^2}\cos(q x+ \theta),\\
&v^{(0,1)}=0.
\end{aligned}
\ee
where $E_{0x}=E_0 \cos \alpha,E_{0y}=E_0 \sin \alpha$.Substituting  Eq.~\eqref{1-order_Arb_pol}  into nonlinear terms one can perform high order iteration   and do the next iterations  by using  the following matrix equation:
\be
 \begin{pmatrix}
    n \\  v_x \\ v_y 
\end{pmatrix}^{(i,j)}=\frac{-i}{s^2 \hat{q}^2-\hat{\omega}(i\gamma+\hat{\omega})}\begin{pmatrix}
i\gamma+\hat{\omega} & \hat{q} & 0 \\ \hat{q} s^2 & \hat{\omega} & 0\\ 
0 & 0 & \hat{\omega}-\frac{\hat{q}^2 s^2}{i\gamma+\hat{\omega}}
\end{pmatrix}~\begin{pmatrix}
  J_n \\ J_{vx} \\ J_{vy}  
\end{pmatrix}^{(i,j)},  
 \ee
where   $\hat{q}= -i\partial_x$, $\hat{\omega}= -i\partial_t$ and  
\begin{equation}
 \begin{pmatrix}
  J_n \\ J_{vx} \\ J_{vy}   
\end{pmatrix}^{(i,j)}=\begin{pmatrix}
  - \partial_x (n v_x) \\ -v_x \partial_x v_x\left(1+2/\Gamma\right)-(2/ \Gamma) v_y\partial_x v_y \\
  - v_x\partial_x v_y
\end{pmatrix}^{(i,j)},   
\end{equation}
are non-linear sources arising  in the $(i,j)$ order of the perturbation expansion.  In order to calculate $n^{(2,0)}$ and $n^{(1,1)},$ we need to calculate   
\begin{equation}
    \left(\begin{array}{c} J_{n} \\ J_{vx} \\ J_{vy} \end{array}\right)^{(2,0)}=    \left(\begin{array}{c} -\partial_{x} (v_x^{(1,0)}n^{(1,0)}) \\  -v^{(1,0)} \partial_{x}v^{(1,0)}(1+2/\Gamma) -(2/\Gamma)v_y^{(1,0)}\partial_x v_y^{(1,0)} \\ v_x^{(1,0)}\partial_x v_y^{(1,0)}
    \end{array}\right),
\end{equation}
and 
\begin{equation}
    \left(\begin{array}{c} J_{n} \\ J_{vx} \\J_{vy} \end{array}\right)^{(1,1)}=  \left(\begin{array}{c} -\partial_{x} (v_x^{(1,0)}n^{(0,1)}) \\  0 \\ 0 \end{array}\right),
\end{equation}
where $n^{(1,0)},n^{(0,1)},v^{(1,0)}$ are given by Eq.~\eqref{1-order_Arb_pol}. After some algebra, we get  
\be
\label{n20-n11_Arb_pol}
\begin{aligned}
   &  n^{(2,0)}= n^{(2,0)}(x)=\frac{e^2 |E_0|^2 h \cos{(q x)}}{4 m^2 s^2 \Gamma}\left(\frac{(\gamma\Gamma-i\omega(2+\Gamma))\cos^2\alpha}{(\gamma+i\omega)(i\gamma\omega+\omega^2-\omega_q^2)}-\frac{2 \sin^2\alpha}{\gamma^2+\omega^2}\right)+h.c.,
 \\
    & n^{(1,1)}=n^{(1,1)}(x,t) =-\frac{e^2 E_0 U_0 q \cos\alpha\sin{(q x+\theta)}}{2 s^2 m^2 (\omega_q^2-\omega(i\gamma+\omega))}e^{-i\omega t}+h.c..
 \end{aligned}
 \ee
We will find the DC current $\mathbf{J}_{\rm{rat}} $ by the following formula
\begin{equation}
\label{Jvec}
    \mathbf{J}_{\rm{rat}}=\frac{e\tau}{m}N_{0}\left\langle n^{(1,1)}\mathbf{E}-n^{(2,0)}\frac{\partial U}{\partial x}\mathbf{e_x}\right\rangle_{t,x}
\end{equation}
Substituting $n^{(1,1)}$ and $n^{(2,0)}$ into Eq.~\eqref{Jvec}, we get $J_{\rm{rat,x}}$ and $J_{\rm{rat,y}}$: 
 
\begin{equation}
     \frac{J_{\rm{rat,x}}}{J_0}
=\frac{P_0[2\omega^4-3\omega^2\omega_q^2+\omega_q^4+\gamma(2\omega^2+\Gamma\omega_q^2)]+P_{\rm L1}~\omega_q^2(\gamma^2\Gamma+\omega^2-\omega_q^2))}{\tau^2\Gamma(\gamma^2+\omega^2)(\gamma^2\omega^2+(\omega^2-\omega_q^2)^2}
\end{equation}

% in Далее, расчет для постоянного поля $J_dc$ в полной мере повторяет шаги аппендикса А. В результате мы получаем для $x$ и $y$ компоненты постоянного тока следующие резлуьтаты:
% % \begin{equation}
%   J_{{\rm rat},x}=\frac{e^3 \tau N_0  q h|E_0|^2 U_0 \sin{\phi}(\omega^4-\omega^2\omega_{q}^2+\gamma^2(\omega^2+\Gamma\omega_q^2))}{4 m^3 s^2 \Gamma(\gamma^2+\omega^2)(\gamma^2\omega^2+(\omega^2-\omega_q^2)^2)}(1+P_{L1}),
% \end{equation}
\begin{equation}
\frac{J_{{\rm rat},y}}{J_0}
 =\frac{(\omega_q^2-\omega^2)P_{L2}+\gamma\omega P_{c}}{2\tau^2 (\gamma^2\omega^2+(\omega^2-\omega_q^2)^2)}.
\end{equation}
 % $P_0,P_{L1},P_{L2},P_{C}$-параметры Стокса, которые имеют следующий вид:$P_0=1$,$P_{L1}=\cos{2 \alpha}$,$P_{L2}=\sin{2\alpha}\cos{\theta}$,$P_{C}=\sin{2\alpha}\sin{\theta}$ 

  %\end{widetext}
\section{ Viscosity contribution}\label{App-vis}
 In this Appendix, we  solve Eq.~\eqref{vveta}  together with Eq.~\eqref{np} (with $v_x=v$). 
The calcaulations are fully analogous to  the ones presented in  Appendix \ref{App-lin}. 
Here, we present  most important modifications without going into technical details.  We start with calculation of the linear response: $n^{(1,0)},~n^{(0,1)},~v^{(1,0)},~ v^{(0,1)}$. Linearizing Eqs.~\eqref{np} and ~\eqref{vveta}   we find:
%response to  the linear perturbation %$a_x$:
\be
\label{1-order_vis}
  \begin{aligned}
  &  n^{(1,0)}=-\frac{e h E_{0} q\sin(q x)}{2m(\omega_q^2-\omega^2-i\gamma \omega)-i q^2 \omega \eta (\omega)) }e^{-i\omega t}+h.c.,
\\
   & v^{(1,0)}=-\frac{e E_{0} }{2m(\gamma-i\omega)}e^{-i\omega t}-\frac{ e h E_{0} \omega\cos(q x)}{2im(\omega_q^2-\omega(\omega+i\gamma)-i q^2 \omega \eta(\omega) )}e^{-i\omega t}+h.c.,
\\
 & n^{(0,1)}=\frac{e V_0 }{m s^2}\cos(q x+ \theta),\\
&v^{(0,1)}=0.
\end{aligned}
\ee
Substituting  Eq.~\eqref{1-order_Arb_pol}  into nonlinear terms one can perform high order iteration   and do the next iterations  by using  the following matrix equation:
\be
 \begin{pmatrix}
    n \\  v  
\end{pmatrix}^{(i,j)}=\frac{-i}{s^2 \hat{q}^2-\hat{\omega}(i\gamma+\hat{\omega})-i\hat{q}^2\hat{\omega}\hat{\eta}}\begin{pmatrix}
i\gamma+\hat{\omega}+i \hat{q}^2\hat{\omega}\hat{\eta} & \hat{q}  \\ \hat{q} s^2 & \hat{\omega} \\ 
\end{pmatrix}~\begin{pmatrix}
  J_n \\ J_{v}   
\end{pmatrix}^{(i,j)},  
 \ee
where   $\hat{q}= -i\partial_x$, $\hat{\omega}= -i\partial_t$, $\hat{\eta}=\eta/(1-i\hat \omega\tau_{ee})$ , $J_n$ and $J_v$ are determined by Eq.~\eqref{JvJn}.Next, we repeat the steps described in appendix \eqref{App-lin}, and get $n^{(1,1)}$ and $n^{(2,0)}$:
\be
\label{n20-n11_vis}
\begin{aligned}
   &  n^{(2,0)}= n^{(2,0)}(x)=\frac{e^2 |E_0|^2 h \cos{(q x)}(-\gamma\Gamma+i(2+\Gamma)\omega-q^2\Gamma\eta(0))}{4 m^2 s^2 \Gamma(\gamma+i\omega)(\omega_q^2-(i\gamma+\omega)\omega-iq^2\omega\eta(\omega))}+h.c.,
 \\
    & n^{(1,1)}=n^{(1,1)}(x,t) =-\frac{e^2 E_0 U_0 q \sin{(q x+\theta)}(i\gamma+\omega+i q^2 \eta(\omega))}{2 s^2 m^2 (i\gamma+\omega)(\omega_q^2-\omega(i\gamma+\omega)-iq^2\omega\eta(\omega))}e^{-i\omega t}+h.c..
 \end{aligned}
 \ee
% Calculations yield the following equation for two contributions to the current
%\begin{align}
  %  & \frac{J_{{\rm rat},x}^{\rm (I)}}{J_0}= \frac{(\gamma\Gamma-i(2+\Gamma)\omega+q^2\Gamma\eta(0))}{\tau^2 \Gamma(\omega-i\gamma)(i\omega_q^2+(\gamma-i\omega)\omega+q^2 \omega\eta(\omega))}+h.c.,
  %  \\ 
  %  & \frac{J_{{\rm rat},x}^{\rm (II)}}{J_0}=\frac{ (\gamma-i\omega+q^2\eta(\omega))}{\tau^2(-\omega-i\gamma)(i\omega_q^2+(\gamma-i\omega)\omega+q^2 \omega\eta(\omega))}+h.c..
%\end{align}
Substituting $n^{(1,1)}$ and $n^{(2,0)}$ into Eq.~\eqref{Jratn}, we find total current  for viscosity with arbitrary dispersion $\eta(\omega)$:

\begin{equation}
    \frac{J_{{\rm rat},x}}{J_0}=\frac{(\gamma-i\omega)(2\gamma\Gamma-2i\omega+q^2\Gamma\eta(0))+q^2\Gamma(\gamma+i\omega)\eta(\omega) }{\tau^2 \Gamma(\gamma^2+\omega^2)(\omega_q^2-(i\gamma+\omega)\omega-iq^2\omega\eta(\omega))}+h.c.
\end{equation}
We assume that viscosity $\eta(\omega)$  has the following dispersion  Ref.~\cite{Alekseev2018}:
\begin{equation}
\eta(\omega)=\frac{\eta}{1-i\omega\tau_{ee}}, 
\end{equation}
where $\tau_{ee}$ is the electron-electron collision time. Then, the dc photoresponse is given by
\be
\begin{aligned}
  & \frac{J_{\rm{rat,x}}}{J_0}=\frac{1}{\tau^2\Gamma}\frac{2(1+\tau_{ee}^2\omega^2)(\omega^4-\omega^2\omega_q^2+\gamma(\omega^2+\Gamma\omega_q^2))}{(\gamma^2+\omega^2)((1+\tau_{ee}^2\omega^2)(\gamma^2\omega^2+(\omega^2-\omega_q^2)^2))+2q^2\eta\omega^2(\gamma+\tau_{ee}(\omega_q^2-\omega^2))}+\\
 & +\frac{1}{\tau^2\Gamma}\frac{q^2\eta\left[\omega^2\gamma(2+\gamma\Gamma\tau_{ee})+\omega^4 \tau_{ee}(\Gamma-2)+\Gamma\omega_q^2(2\gamma+\tau_{ee}\omega^2(\gamma\tau_{ee}-1))\right]}{(\gamma^2+\omega^2)((1+\tau_{ee}^2\omega^2)(\gamma^2\omega^2+(\omega^2-\omega_q^2)^2))+2q^2\eta\omega^2(\gamma+\tau_{ee}(\omega_q^2-\omega^2))}\\
\label{J-eta}
\end{aligned}
\ee
\section{Directed plasmons} \label{App: directed pl}
Oscillating contributions to the concentration    $n^{(1,1)},$ and $n^{(1,0)}$  calculated in the previous section can be easily presented as:
\begin{align}
    & n^{(1,0)}=A_1\cos(\omega t)\sin(q x)+B_1\sin(\omega t)\sin(q x), \\
    & n^{(1,1)}=A_2\cos(\omega t)\sin(q x+\phi)+B_2\sin(\omega t)\sin(q x+\phi),
\end{align}
where $A_1,B_1,A_2,B_2$ are given by:
\be
\label{A1A2}
\begin{aligned}
    & A_1=\frac{e h E_0 q (\omega^2-\omega_q^2)}{m(\gamma^2\omega^2+(\omega^2-\omega_q^2)^2+q^2\eta\omega^2(2\gamma+q^2\eta))},\\
    & B_1=-\frac{e h E_0 q \omega (\gamma+q^2\eta)}{m(\gamma^2\omega^2+(\omega^2-\omega_q^2)^2+q^2\eta\omega^2(2\gamma+q^2\eta))},\\
    & A_2=\frac{e^2 q E_0 U_0\left[(\gamma^2+\omega^2)(\omega^2-\omega_q^2)+q^2\gamma\eta(2\omega^2-\omega_q^2)+q^4\eta^2\omega^2\right]}{s^2 m^2 (\gamma^2+\omega^2)((\omega^2-\omega_q^2)^2+\gamma^2\omega^2+q^2\eta\omega^2(2\gamma+q^2\eta))},\\
    & B_2=-\frac{e^2 q E_0 U_0\omega\left[\gamma(\gamma^2+\omega^2)+q^2\eta(\omega_q^2+2\gamma^2)+q^4\eta^2\gamma\right]}{s^2 m^2 (\gamma^2+\omega^2)((\omega^2-\omega_q^2)^2+\gamma^2\omega^2+q^2\eta\omega^2(2\gamma+q^2\eta))}.
\end{aligned}
\ee
Here $\eta=\eta(0).$ The amplitudes $C_1$ and  $C_2$ reads:
\begin{align}
  & C_1= \frac{1}{2}\sqrt{(B_1+B_2\cos \phi+A_2\sin\phi)^2+(-B_2\sin\phi+A_2\cos \phi+A_1)^2},\\
   & C_2= \frac{1}{2}\sqrt{(-B_1-B_2\cos \phi+A_2\sin\phi)^2+(B_2\sin\phi+A_2\cos \phi+A_1)^2}
\end{align}
The flux of travelling directed plasmons  is expressed through these coefficients as follows:
\begin{equation}
\label{Jpl}
    S_{\rm{pl}}=s(C_1^2-C_2^2)=s(A_2 B_1-A_1 B_2)\sin\phi
\end{equation}
Substitutin  Eq. ~\eqref{A1A2} into  Eq.~\eqref{Jpl}, and keeping linear in $\eta$ term only, we get 
\begin{equation}
S_{\rm{pl}}=-\frac{e^3 q^4 |E_0|^2 U_0 \omega\eta\sin\phi}{s m^3 (\gamma^2+\omega^2)((\omega^2-\omega_q^2)^2+\gamma^2\omega^2)}.
\end{equation}
Finally, by using dimensionless variables  we arrive at  Eq.~\eqref{Spl} of the main text.
\end{widetext}

\newpage
\bibliography{main.bib}

\end{document}